\documentclass[aps,prb,showpacs,twocolumn,amsmath,amssymb,superscriptaddress,floatfix,nofootinbib]{revtex4-2}
\usepackage[dvipsnames,pdftex,usenames]{color} %options n\'{e}cessaires pour compiler avec pdfLaTex
\usepackage[pdftex]{graphics}
\usepackage{graphicx}
\usepackage{xspace}
\usepackage{bm}
\usepackage{amssymb}
\usepackage{textcomp,amsmath}
\usepackage{physics}
\usepackage{float}
\usepackage{braket}
\usepackage[normalem]{ulem}
\usepackage[utf8]{inputenc}

\usepackage[dvipsnames,pdftex,usenames]{color} %options n\'{e}cessaires pour compiler avec pdfLaTex
\usepackage[pdftex,
            bookmarks=true,
            bookmarksnumbered=false,
            bookmarksopen=false,
            colorlinks=true,
            linkcolor=OK-DarkBlue,
            urlcolor=Blue, %Blue or black,
            citecolor=Blue, %Blue or black
            pdfstartview=FitH,
            pdfstartpage=3,
            hyperfootnotes=true,
            pdfhighlight = /N,
            pdfauthor={Olivier Krebs},
            pdfpagemode=UseNone,
            pdftitle={Coherent spectroscopy of a single Mn-doped InGaAs quantum dot}]
            {hyperref}

\definecolor{OK-DarkBlue}{rgb}{0,0.07,0.4}

\def\xp{$X^+$\xspace}

\def\x0{$X^0$\xspace}
\def\a0{$A^0$\xspace}

\def\da0{$|\!-\!1\rangle$\xspace}
\def\ua0{$|\!+\!1\rangle$\xspace}

\def\muW{~\textmu W\xspace}

\def\mueV{~\textmu eV\xspace}
\def\C2N{Centre for Nanosciences and Nanotechnologies, CNRS, Universit\'{e} Paris-Saclay, UMR 9001,  91120 Palaiseau, France}
\def\IIT{Department of Electrical Engineering, Indian Institute of Technology Kanpur, Kanpur-208016, UP, India}

\begin{document}

% Use the \preprint command to place your local institutional report
% number in the upper righthand corner of the title page in preprint mode.
% Multiple \preprint commands are allowed.
% Use the 'preprintnumbers' class option to override journal defaults
% to display numbers if necessary
\preprint{}

%Title of paper

%\title{Electron and nuclear spin orientation in Mn-doped InAs/GaAs quantum dots : exchange vs hyperfine interaction}
\title{Coherent spectroscopy of a single Mn-doped InGaAs quantum dot}
% repeat the \author .. \affiliation  etc. as needed
% \email, \thanks, \homepage, \altaffiliation all apply to the current
% author. Explanatory text should go in the []'s, actual e-mail
% address or url should go in the {}'s for \email and \homepage.
% Please use the appropriate macro foreach each type of information

% \affiliation command applies to all authors since the last
% \affiliation command. The \affiliation command should follow the
% other information
% \affiliation can be followed by \email, \homepage, \thanks as well.

\author{J. Filipovic}\affiliation{\C2N}
\author{A. Kundu}\author{N. K. Vij}\affiliation{\IIT}
\author{S. F\'{e}cherolle}\author{A. Lema\^itre}\affiliation{\C2N}
\author{S. Gupta}\affiliation{\IIT}
\author{O. Krebs}\email[]{olivier.krebs@cnrs.fr}\affiliation{\C2N}

%\homepage[]{Your web page}
%\thanks{}
%\altaffiliation{}

%Collaboration name if desired (requires use of superscriptaddress
%option in \documentclass). \noaffiliation is required (may also be
%used with the \author command).
%\collaboration can be followed by \email, \homepage, \thanks as well.
%\collaboration{}
%\noaffiliation

\date{\today}

\newcommand{\noteSG}[1]{\textcolor{red}{#1}}
\begin{abstract}

Doping a self-assembled InGaAs/GaAs quantum dot (QD) with a single Mn atom, a magnetic acceptor impurity,  provides a quantum system with discrete  energy levels and original spin-dependent optical selection rules, which thus has large potential  in quantum photonics, notably  as a source of multi-entangled photons. To  investigate this potential further,  we perform   coherent  optical spectroscopy under continuous wave excitation of  the 3-level V-like system formed in such a Mn-doped QD when charged by a single hole. In spite of a large inhomogeneous broadening of the optical transitions, we demonstrate Autler-Townes splitting both by resonant Raman scattering and by probe absorption  spectroscopy for different saturation powers. Analysing these data with a comprehensive model based on optical Bloch equations, we show  evidence for quantum interference within the V-like system and assess the pure dephasing rate between the corresponding spin states.

\end{abstract}

\pacs{78.67.Hc,75.50.Pp,71.70.Gm,76.70.Fz}

%78.67.Hc Optical properties of low-dimensional, mesoscopic, and nanoscale materials and structures  : Quantum dots
%78.55.Cr	Photoluminescence, properties and materials : III-V semicondcutors
%71.35.Pq 	Excitons and related phenomena : 	Charged excitons (trions)
%33.80.Be 	Photon interactions with molecules (see also 42.50.-p Quantum optics) : Level crossing and optical pumping
%75.50.Pp 	Magnetic semiconductors

% insert suggested keywords - APS authors don't need to do this
%\keywords{}

%\maketitle must follow title, authors, abstract, \pacs, and \keywords
\maketitle

% body of paper here - Use proper section commands
% References should be done using the \cite, \ref, and \label commands
%\section{}
% Put \label in argument of \section for cross-referencing
%\section{\label{}}
%\subsection{}
%\subsubsection{}

% If in two-column mode, this environment will change to single-column
% format so that long equations can be displayed. Use
% sparingly.
%\begin{widetext}
% put long equation here
%\end{widetext}

% figures should be put into the text as floats.
% Use the graphics or graphicx packages (distributed with LaTeX2e)
% and the \includegraphics macro defined in those packages.
% See the LaTeX Graphics Companion by Michel Goosens, Sebastian Rahtz,
% and Frank Mittelbach for instance.
%

\begin{figure}[ht!]
\includegraphics[width=0.48\textwidth]{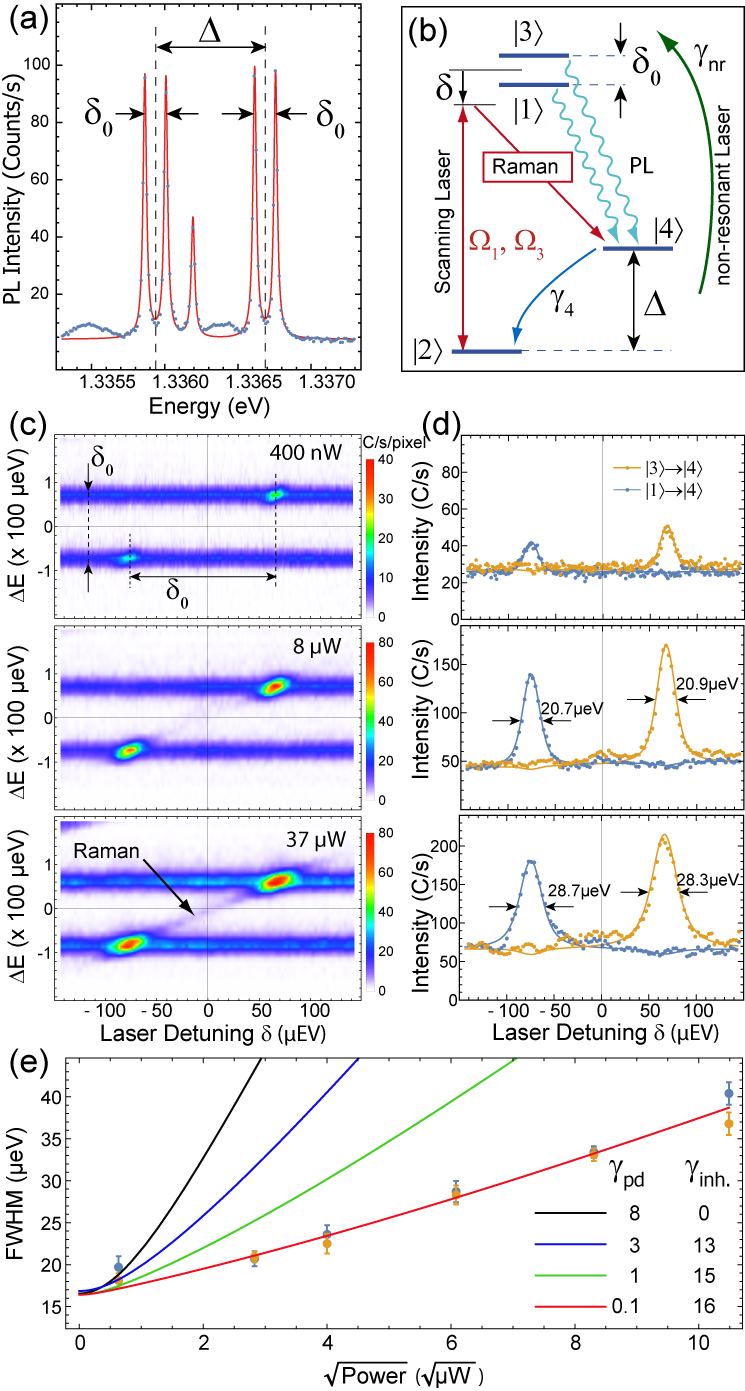}
\caption{(a)  PL spectrum of a Mn-doped InGaAs QD in \xp state measured  under non-resonant excitation  (T=1.5~K).  (b) Schematics  of the implemented resonant absorption spectroscopy. A resonant laser drives the transitions from level $|2\rangle$, and absorption is detected through PL signal recorded at lower energy via spontaneous emission toward level $|4\rangle$. % All levels are twice spin-degenerate in zero magnetic field.
(c) PL spectra as a function of the resonant laser detuning $\delta$ for different incident powers.
%The raw  pixelated  images have been interpolated for a better rendering.
(d) Integrated intensity of both transitions to level $|4\rangle$ as a function of $\delta$. Experiments (points) and model (solid lines). (e)  FWHM of the resonances as a function of the square-rooted power (points), compared with a TLS model (solid lines) for  different couples of homogeneous pure dephasing rate  $\gamma_\text{pd}$ and  Gaussian inhomogeneous broadening $\gamma_\text{inh}$ adjusted to generate $\approx$17\mueV FWHM at low power.}
\label{Fig1}
\end{figure}

\begin{figure}
%\vspace{5mm}
\includegraphics[width=0.48\textwidth]{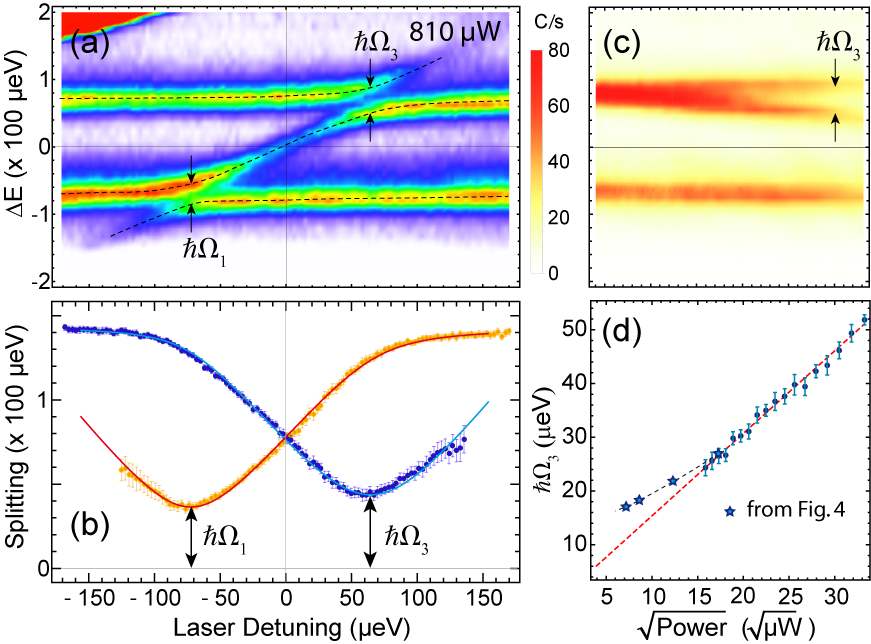}
\caption{(a)  PL spectrum  as a function of the resonant laser detuning $\delta$ at 810\muW incident power, evidencing anti-crossings of the native transitions with the  Raman line (Autler-Townes effect). % \noteSG{[Are the dashed lines from the model?]} OK: No they are only a guide to the eye.
(b) Splitting of the anti-crossing transitions as a function of $\delta$. Experiments (points) and OBE model  (solid lines), enabling us to determine the Rabi frequencys $\Omega_1$ and $\Omega_3$. (c)  PL spectrum as a function of incident power $P_\text{in}$ for a resonant excitation of transition $|2\rangle\leftrightarrow|3\rangle$. (d) Power dependence of the Rabi frequency $\Omega_3$ extracted from (c) (points) along with a linear fit $\propto \sqrt{P_\text{in}}$ (red dashed line). Four additional points (stars) come from the probe absorption spectroscopy shown in Fig.~\ref{Fig4}. }
\label{Fig2}
\end{figure}

\begin{figure}
\includegraphics[width=0.47\textwidth]{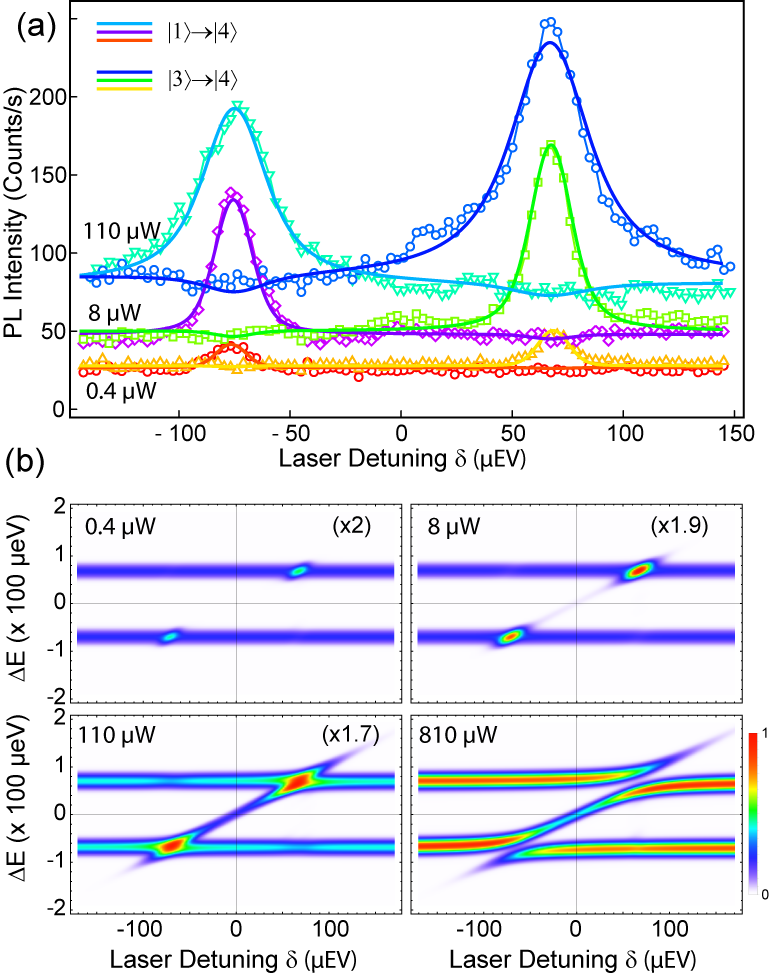}
\caption{(a)  Experimental (symbols) and theoretical (thick solid lines) integrated intensities of both transitions to level $|4\rangle$ for 3 different powers (0.4, 8 and 110 \muW).  (b) Simulated PL spectra as a function of the resonant laser detuning $\delta$ for different incident powers. A scaling factor is applied as indicated for the different powers with respect to the color scale  used for 810\muW.}
\label{Fig3}
\end{figure}

\begin{figure}
\includegraphics[width=0.48\textwidth]{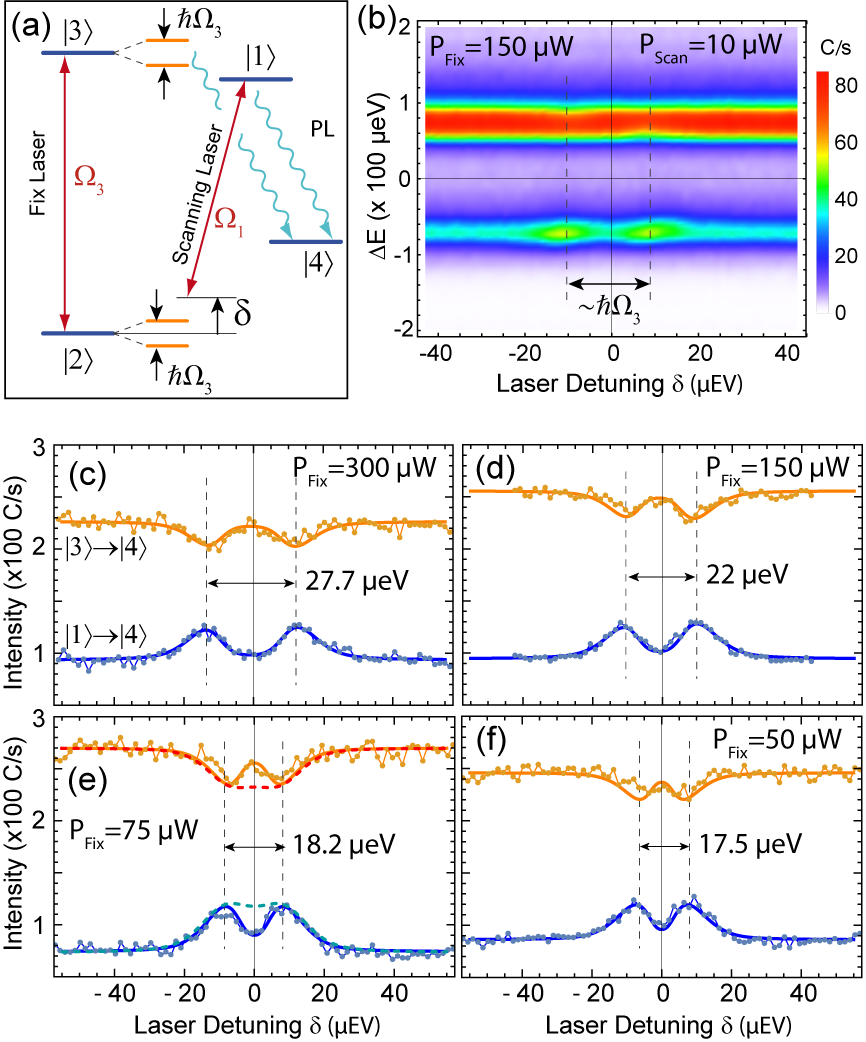}
\caption{ (a)  Schematics of 2-laser spectroscopy. The $|2\rangle-|3\rangle$ transition is resonantly driven by a fixed laser with Rabi frequency $\Omega_3\propto \sqrt{P_3}$, while a second laser of Rabi frequency $\Omega_1\propto\sqrt{P_1}$ scans the $|2\rangle-|1\rangle$ transition as a function of its detuning $\delta$. The levels split  by $\hbar\Omega_3$ (in orange) represent the photon-dressed states in the rotating frame.  (b) PL spectra emitted towards level $|4\rangle$ as a function of $\delta$ for excitation powers $P_3=150$\muW and $P_1=10$\muW. (c)-(f) Integrated PL intensities of the $|1\rangle-|4\rangle$  (symbols, blue) and  $|3\rangle-|4\rangle$ (symbols, orange) lines, for different power settings, along with fits of the model (thick solid lines). In (e) a simulations with large pure dephasing $\gamma_\text{pd}=5$\mueV is also shown (dashed lines).}
\label{Fig4}
\end{figure}

\begin{figure}
\includegraphics[width=0.48\textwidth]{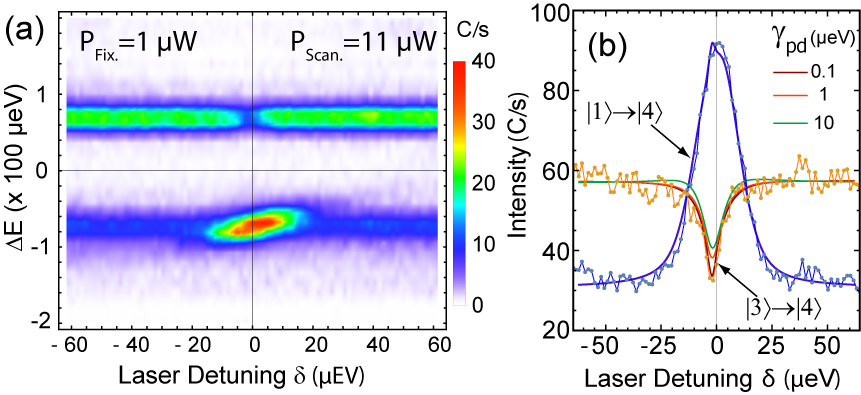}
\caption{(a) PL spectra emitted towards level $|4\rangle$ as a function of detuning $\delta$ of the scanning laser at $P_1=11$\muW in presence of a weaker resonant laser at $P_3=1$\muW, and    (b) corresponding integrated intensities along with a theoretical simulation for different pure dephasing $\gamma_\text{pd}$, see text.}
\label{Fig5}
\end{figure}

Semiconductor quantum dots (QDs) doped by a single  magnetic dopant, namely an impurity  carrying a localized electronic spin, have been studied in the last two decades in order to investigate the exchange interaction with the  spin of a  confined electron or hole,  in the few-spin regime~\cite{Besombes-PRL04,Leger2005,Leger2006,Leger2007,Kudelski2007,Krebs2009,Trojnar2011,Trojnar2013,Mendes2013}, and to assess the potential of such a system as a solid-state quantum bit~\cite{LeGall2009,LeGall2010,LeGall2011,Baudin-PRL11,Goryca2009,Reiter2009,Reiter2011,Besombes2012,Krebs2013,Kobak2014}. This second motivation is still of large interest in the perspective to achieve an ideal spin-based quantum memory coupled to flying qubits like photons. This is notably true in association with  optically active III-V QDs embedded in an optical cavity which are likely the most advanced deterministic sources of indistinguishable single photons~\cite{Somaschi2015,zhai2020low,tomm2021bright}, while being  promising  for the generation of multi-entangled photons in a cluster state. In this respect, the Lindner-Rudolph (LR)  protocol~\cite{Lindner-Rudolph2009} aiming at generating such light states has been successfully implemented with undoped InGaAs/GaAs QDs by exploiting the possibility to entangle the polarization degree of freedom of photon emitted by the  QD with  the spin of a single resident charge~\cite{coste2023high,cogan2023deterministic} or dark exciton~\cite{schwartz2016deter}. However the  scaling up of the LR protocol towards highly entangled photonic cluster states  turns out to be   limited by imperfections of the so far  realized spin-photon interface\cite{Tiurev2021}.\\
\indent In this context, the specific optical selection rules of an  InGaAs/GaAs QD doped by a single Mn atom~\cite{Baudin-PRL11} could be exploited following  a  protocol which has been recently proposed for the generation of cluster states with a single hole spin hosted in a QD molecule~\cite{vezvaee2022deter}.  Indeed,  both types of systems exhibit under specific conditions quite similar optical transitions  involving  4 levels and enabling both the optical  manipulation and readout of a qubit in the ground state. On the path towards this challenging objective, our present work aims to investigate continuous wave (cw) coherent  spectroscopy~\cite{Xu2007,chen2017polar,vora2015spin} of a self-assembled InGaAs/GaAs QD doped with a single Mn atom  to assess coherence between different spin states associated with this  magnetic dopant. In spite of a significant inhomogeneous broadening of the optical transitions, we observe Autler-Townes splittings  by driving two-level transitions far above the saturation regime, while  probe absorption experiment based on   V-like optical transitions clearly reveals quantum interference indicating coherence between the excited spin states.\\
\indent The Mn-doped QD used in the present study is from a sample consisting   of  a single layer of InGaAs/GaAs QD's grown by molecular beam epitaxy with  a very low effective concentration of Mn dopants~\cite{Krebs2009}.  Such a Mn-doped QD has to be found through  scanning  the sample area with a micro-photoluminescence (\textmu-PL) set-up, and is selected via its characteristic spectral signature arising from the exchange interaction of confined carriers with  the single Mn impurity, a neutral acceptor (denoted by \a0)  of effective spin $J=1$~\cite{Kudelski2007}. Due to residual p-type doping, the single Mn-doped QD studied here is also positively charged by an additional hole, as evidenced by  its specific magneto-optical signature~\cite{Krebs2009}. Under optical excitation, an additional  electron-hole pair is generated in the QD and a positive trion \xp  (2 holes, 1 electron) is thus created. In this positive trion, the both  holes   are paired in a singlet configuration, a state for which the exchange interaction with \a0 mostly vanishes~\footnote{The exchange interaction of \a0 with the unpaired electron spin was found much smaller than  the \a0 fine structure splitting for this QD and  is thus neglected.}. As a result, the \textmu-PL spectrum of this Mn-doped and positively charged QD, under zero magnetic field, mainly consists of two doublets as shown in Fig.~\ref{Fig1}(a)~\footnote{ A fifth line associated to the \a0 state $J_z=0$ is also visible in Fig.~\ref{Fig1}(a), but did not play any role in the resonant experiments that were performed, hence is not considered further in this work.}. These four spectral peaks correspond to optical transitions from the two \xp-\a0 levels, denoted as $|1\rangle$ and $|3\rangle$ in Fig.~\ref{Fig1}(b), towards the two hole-\a0 levels, denoted as $|2\rangle$ and $|4\rangle$. The former are split by  the \a0 fine structure $\delta_0$, a term resulting from the coupling of \a0 spin states $J_z=\pm1$ due to the QD-induced anisotropy in the plane perpendicular to the QD growth axis $z$, while the latter are  split by the hole-\a0 exchange energy $\Delta$, see Ref.~\onlinecite{Kudelski2007} for details.\\
\indent In zero magnetic field (configuration of the present work), all these levels are twice spin-degenerate. For instance, the level $|2\rangle$ corresponds to the ferromagnetic configuration  of the hole-\a0  spins reading $|\!\Uparrow,+1\rangle$ or  $|\!\Downarrow,-1\rangle$ (in the limit $\Delta\gg\delta_0$), and the level $|4\rangle$ corresponds to the anti-ferromagnetic configuration $|\!\Uparrow,-1\rangle$ or  $|\!\Downarrow,+1\rangle$, where $\Uparrow, \Downarrow$ denote the  eigenstates  for the heavy-hole spin projection along $z$, and $|\pm1\rangle$ refer to the $J_z$ component of \a0 spin. Depending on the hole spin state, the allowed  optical transitions are left- or right-circularly polarized around the optical $z$ axis. However, in the present work where only linearly polarized optical excitation and detection  are considered, the hole spin degree of freedom  remains in a totally mixed state with the same probability to be up or down, so that it can be ignored in the description and analysis of the following experiments. Hence, the Mn-doped QD system can be seen as the  four-level system  represented in Fig.~\ref{Fig1}(b) accommodating  four optical transitions which are active for  any linear polarization orthogonal to $z$ and are of similar oscillator strengths,   as observed in the PL spectrum of Fig.~\ref{Fig1}(a).
\\
\indent We performed the coherent spectroscopy of this Mn-doped QD with a dark-field confocal  \textmu-PL set-up at low temperature ($\leq$1.5~K) similar to the one described in Ref.~\cite{kuhlmann2013dark}. Here, we use a  3.1-mm~focal length aspheric lens ($0.68$~N.A.) for focusing the excitation continuous wave (cw) laser and for collecting the photoluminescence signal. This lens is actuated by piezo-stages and mounted along with the QD sample in a split-coil magneto-optical cryostat.   To  reject  the reflected laser, the confocal set-up uses a cross configuration of linear polarizations  for the excitation and collection paths along with single mode optical fibers~\cite{kuhlmann2013dark}. By finely adjusting the focusing on the collection fiber input  and the angles of a quarter-wave plate and an analyzer for cross polarization, the laser rejection of our set-up reaches at least $10^{-6}$  over the typical 300\mueV  range scanned by our tunable  cw laser around the energy of 1.336~eV, while the coupling efficiency to the collection fiber amounts to about 50\% in parallel polarization configuration. In spite of the  high performance of our dark-field microscope, resonance fluorescence couldn't be detected for the studied Mn-doped QD, as the PL signal, even below saturation, remains a few order of magnitude smaller than the stray light from the reflected laser. There is indeed no resonant cavity structure here to increase the light-matter interaction and to  better  extract photons from the high-index semiconductor host. Moreover, as shown below,  it turned out that the studied QD experiences a large inhomogeneous broadening reducing the effective QD brightness under resonant excitation by a factor $\sim 20$.\\
\indent To circumvent this issue, we  exploit the specific structure of transitions of a Mn-doped QD as illustrated in Fig.~\ref{Fig1}(b) and described below. Our experimental investigations focus on the V-type system made of the transitions $|1\rangle\leftrightarrow|2\rangle$ and $|3\rangle\leftrightarrow|2\rangle$. They are resonantly driven with a cw tunable mode-hop-free laser line detuned by $\delta$ from the central energy of both transitions and with Rabi frequency $\Omega_1$ and $\Omega_3$, respectively, while  the populations of levels $|1\rangle$ and $|3\rangle$ are monitored via their photoluminescence emitted by recombining towards level $|4\rangle$. The typical redshift  $\Delta\simeq 750$\mueV of these transitions to level $|4\rangle$  enables us to spectrally filter out the laser stray light with a 0.6-m~focal length double spectrometer. Its intermediate slit  is  adjusted to  record PL spectra, with a Nitrogen-cooled CCD array camera, limited to a typical 400\mueV range centered at about 1.3367~eV, namely around the $|1\rangle\rightarrow|4\rangle$ and $|3\rangle\rightarrow|4\rangle$ PL lines.\\
\indent Figure~\ref{Fig1}(c)  shows series of such PL spectra as a function of the laser detuning $\delta$ for three different excitation powers (400~nW, 8~\muW and 37~\muW). Both $|1\rangle\leftrightarrow|4\rangle$ and $|3\rangle\leftrightarrow|4\rangle$ PL lines are visible at all detunings because a weak non-resonant excitation laser at 1.95~eV is used to keep the QD charge state with a resident hole\footnote{This laser is strongly defocused because of the chromaticity of the aspheric lens. Its incident power of a few \muW is adjusted to maximise the intensity of the trion lines, while vanishing the neutral exciton ones.}. Resonances are however clearly observed on the intensity of these lines for the detunings $\delta=\pm\delta_0/2$, namely when the scanning laser is at resonance with either the $|1\rangle\leftrightarrow|2\rangle$ or $|3\rangle\leftrightarrow|2\rangle$ transition and increases the populations of the corresponding level $|1\rangle$ or $|3\rangle$ significantly. Out of resonance, the scanning laser  produces a Stokes line  due to  Raman scattering from level $|2\rangle$ to level $|4\rangle$, as observed for QD molecules~\cite{vora2015spin}. This line (denoted as 'Raman' in Fig.~\ref{Fig1}(c))  is visible at high power only  and  follows the laser detuning,  joining the successive resonances of the main PL lines.\\
\indent For a more quantitative analysis, we use the PL spectra (Fig.~\ref{Fig1}(c)) to calculate the spectrally integrated intensity of each  PL  line ($|1\rangle\leftrightarrow|4\rangle$ and $|3\rangle\leftrightarrow|4\rangle$) as a function of the laser detuning, see Fig.~\ref{Fig1}(d). The obtained intensity profiles reveal  surprisingly broad resonances with full width at half maximum (FWHM) of the order of 17\mueV at low power (400~nW) as compared to the expected natural linewidth of 0.8\mueV of InGaAs QDs  due to their radiative lifetime of $\sim$1~ns. Moreover, the FWHM of these resonances does not increase dramatically when the power increases far above the saturation threshold of $\sim$1\muW, which points toward a dominant  non-inhomogeneous  term for this broadening. To further support this point, in Figure ~\ref{Fig1}(e)  we compare the observed power dependence of the resonance FWHMs of the two PL lines  to the predictions of a two-level system (TLS)  model with 1~ns lifetime, including both an additional  homogeneous pure dephasing $\gamma_\text{pd}$~\cite{meystre2007elements}, and an inhomogeneous broadening  described by a Gaussian distribution with FWHM of $\gamma_\text{inh}$. The experimentally obtained FWHMs are represented by solid circles. Solid lines  represent the FWHM of the resulting Voigt profile  obtained from the TLS model for different sets of $\gamma_\text{pd}$ and $\gamma_\text{inh}$  enabling  to reproduce the observed 17\mueV FWHM at low power. From this comparison, it clearly turns out that the moderate power broadening requires weak pure dephasing $\gamma_\text{pd}\ll 1$\mueV  (in energy unit with $\hbar=1$) along with an  inhomogeneous broadening $\gamma_\text{inh} \approx$ 16\mueV.\\
\indent The presence of  inhomogeneous broadening  does not forbid, in principle, the possibility to generate and detect quantum-coherent effects within the studied V-like system. It only hinders experimental investigations in the weak excitation regime. The power dependence of the resonant Raman scattering (RRS) signal was thus further investigated in the regime where the Rabi frequency largely exceeds $\gamma_\text{inh}$. The RRS spectrum generated under 810\muW incident power is shown in Fig.~\ref{Fig2}(a) as a function of laser detuning. In contrast to   Fig.~\ref{Fig1}c, there is no well-marked resonance but a strong Raman line which follows the laser detuning and  anti-crosses with the PL lines $|1\rangle\rightarrow|4\rangle$ and $|3\rangle\rightarrow|4\rangle$. These anti-crossings are the signature of the Autler-Townes splitting (ATS) of these PL transitions as a result of photon-dressing the $|1\rangle\leftrightarrow|2\rangle$ and $|3\rangle\leftrightarrow |2\rangle$ transitions with a strong laser field. As discussed below, this interpretation is fully  supported by an optical Bloch equation (OBE) model  where a single laser field which drives both $|1\rangle\leftrightarrow|2\rangle$ and $|3\rangle\leftrightarrow |2\rangle$ transitions is treated classically and  the results are obtained under the standard rotating wave approximation, see App.~\ref{AppendixA} for details.\\%\noteSG{[Should Fib 3b be referred here?]}
\indent Precisely measuring ATS is of particular interest to deduce the actual Rabi angular frequencies $\Omega_1$ and $\Omega_3$  at a given incident power. With this aim, we first extract the energy positions of the three distinct lines in Fig.~\ref{Fig2}(a) for each laser detuning, and then plot  the energy difference for each pair of  successive lines, see Fig.~\ref{Fig2}(b). From the minimum of these curves we deduce  $\hbar\Omega_1=37\pm2$\mueV and $\hbar\Omega_3=44\pm3$\mueV. By using these values as parameters in our OBE model, we  then reproduce the full evolution of the measured splittings as shown by the solid lines in Fig.~\ref{Fig2}(b). The  power dependence of the Rabi frequency $\Omega_3$ is further evidenced by varying the incident laser power $P_\text{in}$ at  resonance with transition $|3\rangle\leftrightarrow |2\rangle$  ($\delta=+\delta_0/2$), as shown in Fig.~\ref{Fig2}(c). ATS of the $|3\rangle\rightarrow|4\rangle$ PL line shows up for  $P_\text{in}\gtrsim 250$\muW and then follows the expected dependence of the Rabi frequency $\hbar\Omega_3=k_3 \sqrt{P_\text{in}}$ (Fig.~\ref{Fig2}(d)) with $k_3=49~\text{\textmu eV}/\sqrt{\text{mW}}$. Similarly, we obtain for the transition $|1\rangle\leftrightarrow |2\rangle$  $\hbar\Omega_1=k_1 \sqrt{P_\text{in}}$, with $k_1=41~\text{\textmu eV}/\sqrt{\text{mW}}$. Interestingly, these different  coefficients $k_i$ ($i \in \{1,3\}$) indicate a  difference in the oscillator strength of the respective transitions. This notably explains the difference of intensity for the two successive resonances in Fig.~\ref{Fig1}(d), and also implies that the radiative decay rates $\gamma_i$ of level $|i\rangle$ verify the relation $\gamma_1/\gamma_3= (k_1/k_3)^2$, see App.~\ref{AppendixA}.\\
\indent Thanks to the determination of Rabi frequencies from the measured ATS, it is possible to fit with accuracy the remaining parameters of the OBE model in order to reproduce the RRS spectra over the full range of investigated powers, see App.~\ref{AppendixA} for details.  Figure~\ref{Fig3}(a) presents the results of the model fit along with the integrated PL intensities for three different powers. Assuming a typical radiative lifetime  $\tau_r= 800$~ns for level $|3\rangle$ (as usually observed for InGaAs QDs at the same wavelength), the radiative decay rates $\gamma_1$ and $\gamma_3$ are fixed as constants determined by $\gamma_3=\hbar/\tau_r=$0.8\mueV and $\gamma_1=\gamma_3(k_1/k_3)^2=$0.56\mueV. While the relaxation rate $\gamma_4$ of level $|4\rangle$, the pure dephasing $\gamma_\text{pd}$ of levels $|1\rangle$ and $|3\rangle$ are considered as independent of the power, we keep the possibility for the non resonant excitation rate $\gamma_\text{nr}$ to increase with the power, as the driving laser likely gives rise  to additional non-resonant excitation mechanisms at higher incident powers. The PL intensity profiles from levels $|1\rangle$ and $|3\rangle$  provided by OBE model are then convoluted by a Gaussian distribution of 16\mueV FWHM to take into account the inhomogeneous broadening. Fitting the parameters $\gamma_4$ and $\gamma_\text{pd}$ for these experimental  measurements  gives $\gamma_4=0.05\pm0.02$\mueV and $\gamma_\text{pd}=0.2\pm0.1$\mueV. As shown in Fig.~\ref{Fig3}(a) with these parameters, the model  reproduces quite satisfactorily the main experimental features like the relative intensities of  both resonances and their power-broadening.\\
\indent To compare the OBE model with the measured  RRS  spectra as a function of the detuning and for different powers, we also calculated the theoretical spectra by using the Wiener-Khinchin theorem and the quantum regression theorem, see App.~\ref{AppendixB}. Here also the inhomogeneous broadening  is included and assumed to consist of a rigid spectral wandering of both levels $\ket{1}$  and $\ket{3}$ with respect to both levels  $\ket{2}$  and $\ket{4}$. The Raman line is therefore not broadened by this mechanism because the splitting $\Delta$ between the levels  $\ket{2}$  and $\ket{4}$ remains constant. The calculated RRS spectra presented in Fig.~\ref{Fig3}(b) also take into account the finite spectral resolution (18\mueV) of our set-up. The main features of the experimental RRS spectra are well reproduced in these theoretical plots. The Raman line intensity, hardly visible below 10\muW increases progressively with the incident power while the resonances spread out, and the anti-crossing which is a  signature of ATS shows up at very high power only, in agreement with the experimental results. As discussed in App.~\ref{AppendixB}, our OBE model also captures quite well the interference of both Raman fields associated with both the excited levels $|1\rangle$ and $\ket{3}$. Due tot the relative sign of the four involved dipolar moments, this interference turns out to be constructive  for  $|\delta|<\delta_0/2$ (the Raman line is visible)  and destructive for $|\delta|>\delta_0/2$  (the Raman line vanishes).\\
\indent  To access the coherence between the upper states of this V-like system, probe absorption spectroscopy is performed with two distinct laser lines. In such a 2-color experiment, a laser line is set at the resonance of one of the V-like transitions, while a second laser is scanned across the other transition. This is illustrated in Fig.~\ref{Fig4}(a), where a laser at resonance with transition $\ket{2}\leftrightarrow\ket{3}$  generates ATS associated with the Rabi frequency $\Omega_3$ for this transition. The wavelength of a second  tunable laser line is  swept  across the transition $\ket{2}\leftrightarrow\ket{1}$ with  associated Rabi frequency $\Omega_1$. This enables us to probe the absorption spectrum of transition $\ket{2}\leftrightarrow\ket{1}$ in the presence of ATS by monitoring the PL intensity  from level $\ket{1}$ towards level $\ket{4}$. A typical example of PL intensity map is shown in Fig.~\ref{Fig4}(b) for a power of 150 (10)\muW for the fixed (scanning) laser. In this  power regime (where $P_\text{fix} < 300$\muW ), ATS  is not directly resolved on the spectrum of $\ket{3}\rightarrow\ket{4}$ PL line because of our set-up resolution, but  is clearly revealed on the intensity of the $\ket{1}\rightarrow\ket{4}$ PL line which shows two  resonances split by $\sim\hbar\Omega_3$ when the probe laser detuning is scanned.\\
\indent For a more quantitative analysis, integrated intensities of both PL lines are reported as a  function of the probing laser detuning in Fig.\ref{Fig4}(c)-(f) for different powers of the fixed and scanning laser (with $P_\text{scan}< 10 P_\text{fix}$) and compared with our OBE model  adapted to this experimental configuration with two distinct laser fields and still including inhomogeneous broadening, see App.~\ref{AppendixC}.  The simulated intensities  (thick solid lines)  reproduce quite well the main features of these measurements by using the same model parameters as established above for the single-laser spectroscopy, except for $\gamma_\text{nr}$ which is fitted for each measurement. Actually, the ATS due to the fixed laser is evident on  both PL lines, either as a pair of dips ($\ket{3}\rightarrow\ket{4}$, orange line) or a double peak ($\ket{1}\rightarrow\ket{4}$, blue line). As expected the splitting between these dips or peaks  decreases  when $P_\text{fix}$ is reduced. However, it  deviates from the theoretical ATS given by $\hbar\Omega_3=k_3\sqrt{P_\text{fix}}$ at  low  power when $\hbar\Omega_3\lesssim\gamma_\text{inh}$. Indeed, the  peak-to-peak splittings reported as  stars in Fig.\ref{Fig2}(d) saturate above $\sim$16\mueV.   This effect which is well reproduced by our model, is actually a consequence of the inhomogeneous broadening which leads to a continuous evolution from ATS regime to spectral hole burning regime consisting of  a dip in the 16\mueV inhomogeneous linewidth probed by the scanning laser~\cite{meystre2007elements}. Still, the contrast of the observed  double-peak or double-dip, which can be interpreted as a quantum interference between two pathways quenching the absorption of the scanning laser at resonance,  is a sensitive feature which conveys information on the coherence generated between levels $\ket{1}$ and $\ket{3}$. The crucial parameter of the model which determines this contrast is the pure dephasing rate $\gamma_\text{pd}$. Simulations  with $\gamma_\text{pd}=0.2$\mueV (solid lines in Fig.~\ref{Fig4}(c)-(f)) provide a good agreement, while the contrast almost vanishes for  a larger value $\gamma_\text{pd}=5$\mueV  (dashed line in Fig.~\ref{Fig4}(e)).\\
\indent The establishment of a coherence between levels $\ket{1}$ and $\ket{3}$ is also evidenced in a regime where the fixed laser field at resonance with transition $\ket{2}\leftrightarrow\ket{3}$ is much weaker than the scanning laser and  does not produce significant ATS, see Fig.~\ref{Fig5}. In this regime, the fixed laser no longer perturbs the excitation by the scanning laser, and the PL intensity of $\ket{1}\rightarrow\ket{4}$  transition  shows basically  the same resonance profile as in Fig.~\ref{Fig1} and Fig.~\ref{Fig2} with a single laser. In contrast, the PL intensity of  $\ket{3}\rightarrow\ket{4}$ transition shows now a pronounced, narrow ($\sim 9$\mueV) dip whose origin has two contributions. The most obvious one is the depletion of level $\ket{2}$ population when the scanning laser is at resonance, an effect which is reminiscent of bleaching by spectral hole burning. Another contribution  comes  from a destructive quantum interference  associated to the generation of coherence between levels $\ket{1}$ and $\ket{3}$. Indeed, as shown in Fig.~\ref{Fig5}(b) where theoretical curves are plotted for different values of $\gamma_\text{pd}$, reducing the pure dephasing significantly improves the simulation by enhancing  and narrowing  the dip. In this regard, it is interesting to note that the width of this dip is essentially independent on the inhomogeneous broadening in our OBE model, because the dip takes place under the condition that the frequency difference between the two laser matches the level splitting $\delta_0$. It thus depends mostly on the Rabi frequencies of the laser and on the population decay rates and pure dephasing.\\
\indent In summary,  our present work  demonstrates the establishment of coherence within the upper states of a  3-level V-like system provided by an InGaAs QD doped with  a single Mn atom, when driving the optical transitions with two resonant laser fields. The  extracted dephasing rate $\gamma_\text{pd}$  of this coherence is weak as compared to the radiative lifetime but not  negligible. Still, it might actually include some inhomogeneous contribution which is not considered in our model and which could be inherited from the inhomogeneous broadening of the optical transitions themselves. In future works, it would be anyway more interesting to study the case of a neutral exciton in this type of Mn-doped QDs, since it provides at a given longitudinal magnetic field both a 3-level $\Lambda$-like system and  two spin-conserving recycling transitions~\cite{Baudin-PRL11}. In particular, conducting experiments of coherent population trapping would then be possible and could likely give access to longer spin coherence time for the spin states in the ground state manifold. Still, further investigation and application of this type of QDs  remain  drastically conditioned to the possibility of reducing the large inhomogeneous broadening, that we have observed so far in three different samples. This is most likely an issue of material science requiring significant investment in growth, fabrication and characterization.
%\newpage

\begin{acknowledgments}
The authors would like to thank Joel~C.~Sam and Harish~S.~Adsule for fruitful discussions. This work was partially supported by CEFIPRA  (project 64T3-2), the Science and Engineering Research Board (CRG/2021/003265), the French RENATECH network, the General Council of Essonne,  the Region Ile-de-France in the framework of the Domaine de recherche et d'Innovation Majeur (DIM) Quantum Technologies in Paris Region (QuanTiP) program, and  the French National Research Agency (ANR) as part of the ``Investissements d'Avenir'' program (Labex NanoSaclay, reference : ANR-10-LABX-0035).
\end{acknowledgments}

\appendix

\section{Model of optical Bloch equation  for single laser Raman scattering}
\label{AppendixA}

The Mn-doped quantum dot is modeled  as a four-level system as shown in Fig. 1(b). For definiteness, the eigenstates of the four levels $\ket{1}$, $\ket{2}$, $\ket{3}$ and $\ket{4}$  are expressed as a function of the usual eigenstates of the spin projection along the QD growth axis $z$, namely using $\ket{+1/2}$ or $\ket{-1/2}$ for the electron (in trion configuration),  $\ket{+3/2}$ or $\ket{-3/2}$) for the heavy-hole (in the QD ground state), and $\ket{+1}$, $\ket{0}$ or $\ket{-1}$ for the  $A^0$  neutral acceptor considered as a spin 1. The anisotropic strain experienced by $A^0$  in a typical InGaAs QD leads to the following  approximated expression for the four levels \cite{Kudelski2007}:
%\begin{eqnarray}
%\begin{split}
%\ket{2} = \ket{+3/2,+1}\\
%\ket{4} = \ket{+3/2,-1}\\
%\ket{1} = \ket{+1/2} \otimes \frac{\ket{-1} +\ket{+1}}{\sqrt{2}}\\
%\ket{3} = \ket{+1/2} \otimes \frac{\ket{-1} -\ket{+1}}{\sqrt{2}}\\
%\end{split}
%\end{eqnarray}

\begin{eqnarray}
\label{ket}
\begin{split}
\ket{1} &= \ket{\pm1/2} \otimes \left(\alpha\frac{\ket{-1} -\ket{+1}}{\sqrt{2}} + \sqrt{1-\alpha^2}\ket{0}\right)\\
\ket{2} &= \ket{\pm3/2,\pm1}\\
\ket{3} &= \ket{\pm1/2} \otimes \frac{\ket{-1} +\ket{+1}}{\sqrt{2}}\\
\ket{4} &= \ket{\pm3/2,\mp1}\\
\end{split}
\end{eqnarray}
where it is assumed that the hole-$A^0$ exchange $\Delta$ is much larger than the anisotropic splitting term $\delta_0$, and that the electron-$A^0$ exchange is negligible for the studied QD. The parameter $\alpha$  is introduced to reproduce the observed difference of Rabi frequencies ($k_1\neq k_3)$ and  is thus given by $\alpha=k_1/k_3$. % satisfies $0<1-\alpha^2\ll 1$.
Equation~\ref{ket} is mostly useful to properly define the relative phases of the dipolar matrix elements which in turn determine how the different scattered optical fields   interfere when different optical transitions are driven by the same laser field. Each state in Eq.~\ref{ket} is spin degenerate, but the spin state is not kept in labelling the levels $\ket{i}$ ($i\in\{1,2,3,4\}$) because all  the considered optical transitions are spin conserving and determined by a single  QD dipolar matrix element $\mathbf{d}_0 = -e \bra{\pm3/2}\mathbf{\hat{r}} \ket{\pm1/2}$ whereas $ \bra{\pm3/2}\mathbf{\hat{r}} \ket{\mp1/2}=0$.  Taking now into account the $A^0$ part in Eq.~\ref{ket}, we obtain  four dipolar matrix elements for the QD  optical transitions:

\begin{eqnarray}
\label{dipoles}
\begin{split}
    \mathbf{d}_{21} = \frac{-\alpha\mathbf{d}_0}{\sqrt{2}}\hspace{2cm}
    \mathbf{d}_{23} = \frac{\mathbf{d}_0}{\sqrt{2}}\\
    \mathbf{d}_{41} = \frac{\alpha\mathbf{d}_0}{\sqrt{2}}\hspace{2.2cm}
    \mathbf{d}_{43} = \frac{\mathbf{d}_0}{\sqrt{2}}
    \end{split}
\end{eqnarray}

\noindent  These expressions indicate  that the Rabi couplings of a laser optical field of amplitude $\mathbf{E}_0$  and  frequency $\omega_\text{L}$ with  the  $\ket{2}-\ket{i}$   transitions ($i\in\{1,3\}$)   have a phase difference of $\pi$ between them. By introducing the corresponding Rabi frequencies $\Omega_i=-\mathbf{d}_{2i}\cdot\mathbf{E}_0 $ (with energy units assuming $\hbar=1$) and using the rotating wave approximation (RWA),    the Hamiltonian of the four-level system  in the rotated frame of the laser field reads~\cite{meystre2007elements}:

\begin{equation}
\begin{split}
H_\text{sys} =  \Delta \sigma_{44}  - ( \delta
 +\delta_0/2) \sigma_{11} - (\delta - \delta_0/2) \sigma_{33}\\  + \frac{\Omega_{1}}{2}(\sigma_{12}
+ \sigma_{21})+\frac{\Omega_3}{2}(\sigma_{32}+\sigma_{23})
\end{split}
\label{final_h_rot}
\end{equation}

\noindent where the operators $\sigma_{ij}\equiv\ket{i}\bra{j}$ represents the dipole transition operator when $i \neq j$ and the population operator when $i=j$. We set the energy of the ground state level $\ket{2}$ to zero and  denote $\delta$ the laser frequency  detuning with respect to the middle energy of the levels $\ket{1}$ and $\ket{3}$, which are split by  the positive energy $\delta_0$. Resonance conditions for the transitions $\ket{2}\rightarrow\ket{1}$ and $\ket{2}\rightarrow\ket{3}$ corresponds thus to $\delta=-\delta_0/2$ and $\delta=+\delta_0/2$ respectively. Note that the Rabi frequencies $\Omega_1$ and $\Omega_3$ can be set as real numbers but have to be of opposite sign as explained above. The eigenenergies of $H_\text{sys}$ are used to reproduce the experimental splittings reported in Fig.~\ref{Fig2}(c) as a function of the  laser detuning $\delta$, which thus  determines the  values for $|\Omega_1|$ and $|\Omega_3|$ at 810\muW excitation power.\\
\indent The master equation for the density matrix $\rho$  of our system, including the different relaxation channels as well as a non-resonant  excitation term, is described in the Liouvillian formalism as follows:
\begin{equation}
\label{MasterEq}
    \begin{split}
         \displaystyle{\frac{\partial \rho}{\partial t}} & = i[\rho, H_\text{sys}] + \gamma_{12}\mathcal{L}\left(\sigma_{21}\right)\rho + \gamma_{14}\mathcal{L}\left(\sigma_{41}\right)\rho\\
        & +\gamma_{32}\mathcal{L}\left(\sigma_{23}\right)\rho + \gamma_{34}\mathcal{L}\left(\sigma_{43}\right)\rho + \gamma_4\mathcal{L}\left(\sigma_{24}\right)\rho \\
       &  +\gamma_\text{pd}\left[\mathcal{L}\left(\sigma_{11}\right) + \mathcal{L}\left(\sigma_{33}\right)\right]\rho \\
       &  +\frac{\gamma_\text{nr}}{2}\left[\mathcal{L}\left(\sigma_{12}\right)  +\mathcal{L}\left(\sigma_{32}\right) +\mathcal{L}\left(\sigma_{14}\right) + \mathcal{L}\left(\sigma_{34}\right)\right]\rho
    \end{split}
\end{equation}
\noindent where $\mathcal{L}(\hat{O})\rho \equiv \hat{O} \rho \hat{O}^\dagger - \displaystyle{\frac{1}{2}(\hat{O}^\dagger\hat{O} \rho + \rho \hat{O}^\dagger \hat{O})}$ is the Lindblad superoperator, $\gamma_4$ is the  decay rate from level $\ket{4}$ to $\ket{2}$, $\gamma_\text{nr}$ denotes the non-resonant pumping rate from levels $\ket{2}$ and $\ket{4}$, and $\gamma_{i2}, \gamma_{i4}$ are the spontaneous emission decay rates from level $\ket{i}$ to $\ket{2}$ and $\ket{4}$ respectively, for $i\in \{1,3\}$. Interestingly, from the photoluminescence intensity of the different transitions under non-resonant excitation (see Fig.~\ref{Fig1}(a)), it can be concluded  that spontaneous emission rates from the levels $\ket{1}$ and $\ket{3}$ are equally shared between recombination to levels $\ket{2}$ and $\ket{4}$ (within a relative error $<2\%$), which allows us to set  $\gamma_{i4}=\gamma_{i2}\equiv \gamma_{i}/2$. Besides, since $\gamma_{i2} \propto |\mathbf{d}_{i2}|^2$ we can also use Eq.~\ref{dipoles} and set the ratio $\gamma_1/\gamma_3=\alpha^2=(k_1/k_3)^2$.  Equation~\ref{MasterEq}  also includes the important  term describing the pure-dephasing  rate $\gamma_\text{pd}$ for levels $\ket{1}$ and $\ket{3}$. The resulting optical Bloch equations for the density matrix elements $\rho_{ij}=\bra{i}\rho\ket{j}$ which have to be solved in the steady state regime read thus:
\begin{widetext}
\vspace{0.1cm}
\begin{eqnarray}
    \begin{split}
    \frac{\partial \rho_{11}}{\partial t} &= -\gamma_1 \rho_{11}+\frac{\gamma_\text{nr}}{2}(\rho_{2}+\rho_{44}) + i\frac{\Omega_1}{2}(\rho_{12} - \rho_{21})=0\\
     \frac{\partial \rho_{12}}{\partial t} &= \left[-\frac{1}{2}(\gamma_1 +\gamma_\text{nr} + \gamma_\text{pd}) + i(\delta+\delta_0/2)\right]\rho_{12} +i\frac{\Omega_1}{2}(\rho_{11}-\rho_{22}) +i\frac{\Omega_3}{2}\rho_{13}=0\\
     \frac{\partial \rho_{13}}{\partial t} &= \left[-\frac{1}{2}(\gamma_1 +\gamma_3 + 2\gamma_\text{pd}) + i\delta_0\right]\rho_{13} -i\frac{\Omega_1}{2}\rho_{23} +i\frac{\Omega_3}{2}\rho_{12}=0\\
    \frac{\partial \rho_{14}}{\partial t} &= \frac{1}{2}(-\gamma_1 -\gamma_4 -\gamma_\text{nr} - \gamma_\text{pd})\rho_{14} + i\left[\Delta \rho_{14} +(\delta+\delta_0/2)\rho_{14} -\frac{\Omega_1}{2}\rho_{24}\right]=0\\
     \frac{\partial \rho_{22}}{\partial t} &=-\gamma_\text{nr}\rho_{22}+\frac{1}{2}(\gamma_1 \rho_{11}+\gamma_3 \rho_{33})+\gamma_4\rho_{44}+i\frac{\Omega_1}{2}(\rho_{21}-\rho_{12})+i\frac{\Omega_3}{2}(\rho_{23}-\rho_{32})=0 \\
    \frac{\partial \rho_{24}}{\partial t} &=-\frac{\gamma_4}{2}\rho_{24} -\gamma_\text{nr}\rho_{24} + i\left(\Delta \rho_{24} -\frac{\Omega_1}{2}\rho_{14} -\frac{\Omega_3}{2}\rho_{34}\right)=0
    %\frac{\partial \rho_{34}}{\partial t} &= \frac{-1}{2}(\gamma_3 +\gamma_4 + \gamma_\text{nr} + \gamma_\text{pd})\rho_{34} + i(\Delta\rho_{34} + (\delta-\delta_0/2)\rho_{34} - \frac{\Omega_3}{2}\rho_{24})=0
    \end{split}
    \label{OBE6}
\end{eqnarray}
\end{widetext}
Additional equations  involving the level $\ket{3}$ can be deduced from Eq.~\ref{OBE6}  by swapping  the indices 1 and 3, and by replacing $\delta_0$ by $-\delta_0$. We also remind that $\rho_{ij}=\rho_{ji}^\star$.\\
\indent This set of equations is solved analytically with Mathematica$^\circledR$ which provides the exact expressions for the populations  $\rho_{11}$ and $\rho_{33}$ as a function of $\delta$. To reproduce the experimental integrated PL intensities in Fig.~\ref{Fig3}(a), we have to include the inhomogeneous broadening. This is done by a numerical convolution of $\rho_{11}(\delta)$ and $\rho_{33}(\delta)$ with a Gaussian function with  FWHM $\Gamma_\text{inh}=15$\mueV which provides the new functions $\tilde{\rho}_{11}(\delta)$ and $\tilde{\rho}_{33}(\delta)$. The  model parameters are fitted by runing an optimisation algorithm on the experimental data with  the theoretical PL intensities given by $\kappa \gamma_1\tilde{\rho}_{11}(\delta)$ and $\kappa \gamma_3\tilde{\rho}_{33}(\delta)$ where $\kappa$ is a common proportionality factor.

% Using the above equations, Wiener-Khintchin theorem, and the relation between electric field operators and atomic transition operators (i.e., $E^+_{41}(t) \propto \sqrt{\gamma_1/2}  \ \mathbf{d_{41}} \ \sigma_{41}(t)$ and $E^+_{43}(t) \propto \sqrt{\gamma_3/2}\  \mathbf{d_{43}} \ \sigma_{43}(t)$), we determine the spectrum of the system in terms of first-order single-time correlations of the transition operators.
% \begin{widetext}
% \begin{equation}
%     \begin{split}
%       S(\omega)= \lim_{t\to \infty}\frac{1}{\pi} Re \Big\{\int_{-\infty}^{t} ( \frac{\gamma_1}{2}\langle \sigma_{14}(t) \sigma_{41}(t+\tau)\rangle + \frac{\sqrt{\gamma_1\gamma_3}}{2}\sigma_{14}(t) \sigma_{43}(t+\tau)\rangle \\ + \frac{\sqrt{\gamma_3\gamma_1}}{2}\langle \sigma_{34}(t) \sigma_{41}(t+\tau)\rangle + \frac{\gamma_3}{2}\langle \sigma_{34}(t) \sigma_{43}(t+\tau)\rangle ) e^{i\omega\tau}d\tau \Big\}
%     \end{split}
%     \label{spectrum eq.}
% \end{equation}
% \end{widetext}

% Finally, we use the Quantum Regression Theorem to calculate the correlation of the transition operators in terms of the steady-state solutions of the optical Bloch equations (Eq. \eqref{OBE6}) of our system to calculate the spectrum of the system (Eq. \eqref{spectrum eq.}).

\section{Photoluminescence spectrum}
\label{AppendixB}
% \section{Accounting for broadening effects}
To generate the PL spectrum map shown in Fig.~\ref{Fig3}(b) as a function of laser detuning and energy detection, we first invoke the Wiener-Khintchin theorem which relates the spectrum of the emitted field intensity $S(\omega)$ as a function of the Fourier transform of the first-order  time correlation of the detected electric field as follows~\cite{meystre2007elements}:
\begin{equation}
    \begin{split}
      S(\omega)\propto  \Re\int_{0}^{\infty}  \langle E^-(\tau)E^+(0)\rangle e^{-i \omega\tau} d\tau
    \end{split}
    \label{WK}
\end{equation}
In  our case, the detection window is centered around the transitions  $\ket{1}\rightarrow\ket{4}$ and $\ket{3}\rightarrow\ket{4}$, so that the electric field operator $E^+(t)$ at time $t$ to consider is the sum of the two fields emitted from levels $\ket{1}$ and $\ket{3}$ to level $\ket{4}$ which reads:
\begin{equation}
E^+(t)\propto\sum_{i\in\{1,3\}}\sqrt{\gamma_i}\,\bm{\epsilon}_V\cdot \frac{\mathbf{d}_{4i}}{ \|\mathbf{d}_{4i}\|} \sigma_{4i}(t)
\end{equation}

where $\bm{\epsilon}_V$ is the polarization vector used in detection (in our case a linear vertical polarization). By replacing this expression in Eq.~\ref{WK}, we get:
%is the sum of the  atomic transition operators (i.e., $E^+_{41}(t) \propto \sqrt{\gamma_1/2}  \ \mathbf{d_{41}}  \sigma_{41}(t)$ and $E^+_{43}(t) \propto \sqrt{\gamma_3/2}\  \mathbf{d_{43}} \ \sigma_{43}(t)$), to determine the spectrum of the system in terms of first-order single-time correlations of the transition operators.
%\begin{widetext}
%\begin{equation}
%S(\omega)\propto\Re\int_{0}^{\infty}  \Big(\gamma_1\langle \sigma_{14}(0) \sigma_{41}(\tau)\rangle+\gamma_3\langle \sigma_{34}(0) \sigma_{43}(\tau)\rangle +\sqrt{\gamma_1\gamma_3}\langle \sigma_{34}(0) \sigma_{41}(\tau)\rangle+\sqrt{\gamma_1\gamma_3}\langle \sigma_{14}(0) \sigma_{43}(\tau)\rangle\Big) e^{-i \omega\tau}d\tau
%\label{SpEq}
%\end{equation}
%\end{widetext}

% \begin{widetext}
\begin{equation}
\begin{split}
S&(\omega)\propto\Re\int_{0}^{\infty} d\tau \Big(\gamma_1\langle \sigma_{14}(0) \sigma_{41}(\tau)\rangle+\gamma_3\langle \sigma_{34}(0) \sigma_{43}(\tau)\rangle\\
&+\sqrt{\gamma_1\gamma_3}\langle \sigma_{34}(0) \sigma_{41}(\tau)\rangle+\sqrt{\gamma_1\gamma_3}\langle \sigma_{14}(0) \sigma_{43}(\tau)\rangle\Big) e^{-i \omega\tau}
\end{split}
\label{SpEq}
\end{equation}
% \end{widetext}
To calculate this integral, we next use the quantum regression theorem (QRT) to express the time correlations of the transition operators in terms of the steady-state solutions of the optical Bloch equations~\cite{meystre2007elements}.  From the master equation~\ref{MasterEq},  we first  derive the   $3\times3$ matrix $M$ which determines the evolution of the vector $v_4=(\rho_{14},\rho_{24}, \rho_{34})$ according to $dv_4/dt=M\cdot v_4$. The application of QRT provides then the following relations:

%\begin{equation}
%M=\left(
%\begin{array}{ccc}
% -\frac{\gamma_1}{2}-\frac{\gamma_4}{2}-\frac{\gamma_\text{nr}}{2}-\frac{\Gamma_{31}}{2}+i (\Delta -\Delta_1) & -i
%   \Omega_1 & 0 \\
% -i \Omega_1 & -\frac{\gamma_4}{2}-\gamma_\text{nr}+i \Delta  & -i \Omega_3 \\
% 0 & -i \Omega_3 & -\frac{\gamma_3}{2}-\frac{\gamma_4}{2}-\frac{\gamma_\text{nr}}{2}-\frac{\gamma_{31}}{2}+i (\Delta -\Delta_1-\delta_{13}) \\
%\end{array}
%\right)\end{equation}

%By application of the quantum regression theorem (QRT), this  $M$ matrix allows for the calculation of the 4 terms in above Eq.~\ref{WKintensity} simultaneously, thanks to the linearity of the equations (25)-(27) in Neelesh's notes. In basically one step, we get thus the total intensities   $a_j$'s of the spectral line $j$ associated to the eigenvalues $\lambda_j$ of the matrix $M$ for a given detuning $\delta_1$. Below,  more details are given  about this application of QRT and  the associated maths implemented in the Mathematica code. First, the application of QRT (knowing that $\langle \hat{\sigma}_{ij}\rangle = \rho_{ji}$) gives rise to the following equations:

\begin{equation}
\frac{d}{d\tau}\underbrace{\left(
\begin{array}{ccc}
  \langle \hat{\sigma}_{14}(0)\hat{\sigma}_{41}(\tau) \rangle\\
 \langle \hat{\sigma}_{14}(0)\hat{\sigma}_{42}(\tau) \rangle \\
\langle \hat{\sigma}_{14}(0)\hat{\sigma}_{43}(\tau) \rangle \\
\end{array}
\right)}_{\bm{V}_1(\tau)}=M\cdot
\left(
\begin{array}{ccc}
  \langle \hat{\sigma}_{14}(0)\hat{\sigma}_{41}(\tau) \rangle\\
 \langle \hat{\sigma}_{14}(0)\hat{\sigma}_{42}(\tau) \rangle \\
\langle \hat{\sigma}_{14}(0)\hat{\sigma}_{43}(\tau) \rangle \\
\end{array}
\right)\end{equation}

 and

\begin{equation}
\frac{d}{d\tau}\underbrace{\left(
\begin{array}{ccc}
  \langle \hat{\sigma}_{34}(0)\hat{\sigma}_{41}(\tau) \rangle\\
 \langle \hat{\sigma}_{34}(0)\hat{\sigma}_{42}(\tau) \rangle \\
\langle \hat{\sigma}_{34}(0)\hat{\sigma}_{43}(\tau) \rangle \\
\end{array}
\right)}_{\bm{V}_3(\tau)}=M\cdot
\left(
\begin{array}{ccc}
  \langle \hat{\sigma}_{34}(0)\hat{\sigma}_{41}(\tau) \rangle\\
 \langle \hat{\sigma}_{34}(0)\hat{\sigma}_{42}(\tau) \rangle \\
\langle \hat{\sigma}_{34}(0)\hat{\sigma}_{43}(\tau) \rangle \\
\end{array}
\right)\end{equation}

These are two sets of differential linear equations based on the same matrix $M$ which can be diagonalized (because $M$ is symmetrical) such that $D=P^{-1} M P$ where $P$ is the matrix of the eigenvectors, and $D$ is the diagonal matrix made of the complex eigenvalues $\lambda_i$ of $M$,  with $i\in\{1,2,3\}$. The solutions are given by :

\begin{equation}
\begin{split}
\bm{V}_1(\tau)=P e^{D \tau} P^{-1} \bm{V}_1(0)\\
\bm{V}_3(\tau)=P e^{D \tau} P^{-1} \bm{V}_3(0)
\label{QRT-sol}
\end{split}
\end{equation}
Since  $\langle \hat{\sigma}_{ij}(0)\hat{\sigma}_{jk}(0) \rangle=\langle \hat{\sigma}_{ik} \rangle=\rho_{ki}$, the above solution (\ref{QRT-sol}) is then completely defined once  the master equation (\ref{MasterEq}) is  numerically solved in the steady-state regime  (for a given laser detuning $\delta$) and  the  matrix $M$ is numerically diagonalized. Finally the PL intensity spectrum (\ref{SpEq}) is calculated by using the first and third components of the vectors $\bm{V}_1(\tau)$ and  $\bm{V}_3(\tau)$. Doing so, we end up with 4 functions  $f_k(\tau)$ of the form $f_k(\tau)=\sum_{j=1}^{3}a_{kj}\exp(\lambda_j \tau)$  with $k\in\{1,2,3,4\}$, and the Fourier transform of these functions in Eq.~(\ref{SpEq}) can be solved analytically by introducing the real and imaginary parts of the eigenvalues $\lambda_j=-(y_j +i\, x_j)$. The theoretical spectrum $I_\text{PL}(\omega,\delta)$ is then given by :

\begin{equation}
I_\text{PL}(\omega,\delta)\propto\operatorname{Re}\Bigg(\sum_{j=1}^{j=3} \Big(\sum_{k=1}^4 \gamma_k^\text{eff.} a_{kj}\Big) \frac{y_j+i(\omega-x_j)}{(x_j-\omega)2+y_j^2}  \Bigg)
\label{IPL-final}
\end{equation}

where $\gamma_1^\text{eff.}=\gamma_1$, $\gamma_2^\text{eff.}=\gamma_3$, $\gamma_3^\text{eff.}=\gamma_4^\text{eff.}=\sqrt{\gamma_1\gamma_3}$.\\

\indent To generate the PL intensity maps in Fig.~\ref{Fig3}(b),  the function $I_\text{PL}(\omega, \delta)$ is first calculated following Eq.~\ref{IPL-final}, but the energies of the  three spectral lines ($x_j=-\operatorname{Im}\lambda_j$) are shifted by the laser detuning $-\delta$ to come back in the laboratory frame. Then we include both the inhomogeneous broadening (spectral wandering) and the finite spectral resolution of our setup. Like for the  integrated PL intensity in App.~\ref{AppendixA}, we first perform a discrete convolution the function $I_\text{PL}(\omega, \delta)$  with a Gaussian distribution of the effective detuning with  15\mueV FWHM. Since we assume that the inhomogeneous broadening is due to a rigid shift (spectral wandering) of both  levels $\ket{1}$ and $\ket{3}$ with respect to both  levels $\ket{2}$ and $\ket{4}$, a random spectral shift $\delta x$ of the actual laser detuning $\delta$ gives rise to a shift $-\delta x$ of the PL emission energy. This property is  taken into account by performing the convolution on both arguments $\omega$ and $\delta$ of $I_\text{PL}(\omega, \delta)$ in a proper way, so that the energy of the  Raman line does not experience any spectral wandering.  The final step is  a second  convolution of the PL intensity function along the $\omega$ axis by a Gaussian distribution of 18\mueV FWHM corresponding to our spectral resolution.\\

%Finally, we have considered two major broadening effects for the PL spectrum. The first is due to the finite resolution of the detector. The photodetector used in the experiment has a wavelength resolution of around $2\ pm$, resulting in a linewidth broadening of about $18\ \mu eV$. Second is the inhomogeneous broadening arising due to the charge fluctuations in the environment/semiconductor matrix. We estimate the inhomogeneous broadening from low power PL measurements to be $\Gamma_{inh}=16\ \mu eV$, assuming negligible power broadening at these low powers. We incorporate both these broadening effects in our model by convolving the PL spectrum with a Gaussian of full-width half-maximum (FWHM) $=16\ \mu eV$ to capture the effect of inhomogeneous broadening and then convolving again with a Gaussian of FWHM $=18\ \mu eV$ to capture the effect of finite resolution.

%\subsection{Integrated Photoluminescence}

%The integrated PL is evaluated by performing an integration over the frequency space information relevant to the energy levels around the quantum dot energy levels $\ket{1}$ and $\ket{3}$. It can be shown that the Integrated PL, which is essentially the scattered intensity, is effectively a function of the steady-state population.
%\begin{equation}
%    I \propto \langle\sigma_{+}(0) \sigma_{-}(0) \rangle
%\end{equation}
%\\

\section{Model   for two-laser spectroscopy}
\label{AppendixC}
For the probe absorption experiments using  two different lasers,  the time-dependent RWA Hamiltonian describing the interaction with the transitions $\ket{2}\leftrightarrow\ket{1}$ and  $\ket{2}\leftrightarrow\ket{3}$ in the laboratory frame  reads:

\begin{equation}
\begin{split}
H_{\text{sys}} =  \Delta \sigma_{44} + \omega_1 \sigma_{11} + (\omega_1 +\delta_0) \sigma_{33}   \\
+ \left(\frac{\Omega_{1,\text{f}}}{2}\sigma_{21}e^{i \omega_{\text{f}}t} + \frac{\Omega_{3,\text{f}}}{2}\sigma_{23} e^{i \omega_{\text{f}}t} + h.c\right)\\
+ \left(\frac{\Omega_{1,\text{s}}}{2}\sigma_{21}e^{i \omega_{\text{s}}t} + \frac{\Omega_{3,\text{s}}}{2}\sigma_{23} e^{i \omega_{\text{s}}t} + h.c.\right)
\end{split}
\label{two_laser_ham_start}
\end{equation}

\noindent where $\omega_1$ is the frequency of the level $\ket{1}$, $\omega_{\text{f}}$ and $\omega_{\text{s}}$ are the frequencies of the fixed and scanning lasers, and $\Omega_{1,\text{f}}, \Omega_{3,\text{f}}$ and $\Omega_{1,\text{s}},\Omega_{3,\text{s}}$ are their  Rabi frequencies for the transitions associated to levels $\ket{1}$ and  $\ket{3}$, respectively. Since the fixed laser is at resonance with transition $\ket{2}\leftrightarrow\ket{3}$ and the scanning laser frequency $\omega_{\text{s}}$ remains near the frequency of transition $\ket{2}\leftrightarrow\ket{1}$ we transform our Hamiltonian into a rotated frame defined by  $H_0 = \omega_\text{f}\sigma_{33} + \omega_\text{s}\sigma_{11}$  to eliminate the time dependence. The rotated Hamiltonian reads:

\begin{equation}
\begin{split}
H'_\text{sys} &=
%e^{i H_0 t}H'_\text{sys}e^{-i H_0 t}-i e^{i H_0 t} d(e^{-i H_0 t})/dt \\
\Delta \sigma_{44} - \delta_1 \sigma_{11} - \delta_3 \sigma_{33}    \\
&+ \left(\frac{\Omega_{2,\text{f}}}{2}\sigma_{21} e^{i(\delta_3 -\delta_1 + \delta_0)t} + \frac{\Omega_{3,\text{f}}}{2}\sigma_{23} + h.c.\right)\\
&+\left( \frac{\Omega_{1,\text{s}}}{2}\sigma_{21} +\frac{\Omega_{3,\text{s}}}{2} \sigma_{23}e^{-i(\delta_3 -\delta_1 + \delta_0)t} + h.c.\right)
\end{split}
\label{two_laser_ham_rot}
\end{equation}

where $\delta_1 = \omega_{\text{s}}-\omega_1 $ and $\delta_3=\omega_\text{f}-(\omega_1+\delta_0)$.
%and $\delta_3 -\delta_1 + \delta_0 = \omega_\text{fix} - \omega_{\text{scan}}$.
Because of the two laser frequencies, some time-dependent terms  remain in the rotated frame. However, these terms proportional to $ e^{\pm i (\delta_3 -\delta_1 + \delta_0) t}$ can be considered as  fast-rotating with respect to the slow evolution due to  Rabi driving, because  $\delta_3 -\delta_1 + \delta_0 \simeq \delta_0 > \{\Omega_1, \Omega_3\}$, so that in principle they can be dropped  to get the final time-independent two-laser Hamiltonian :

\begin{equation}
\begin{split}
H'_\text{sys} =  \Delta \sigma_{44} - \delta_1 \sigma_{11} - \delta_3 \sigma_{33}  \\
+ \frac{\Omega_{1,\text{s}}}{2}(\sigma_{21} + \sigma_{12}) + \frac{\Omega_{3,\text{f}}}{2}(\sigma_{23} + \sigma_{32})
\end{split}
\label{H2Final}
\end{equation}

\begin{figure}[htbp]
    \centering
    \includegraphics[scale=0.25]{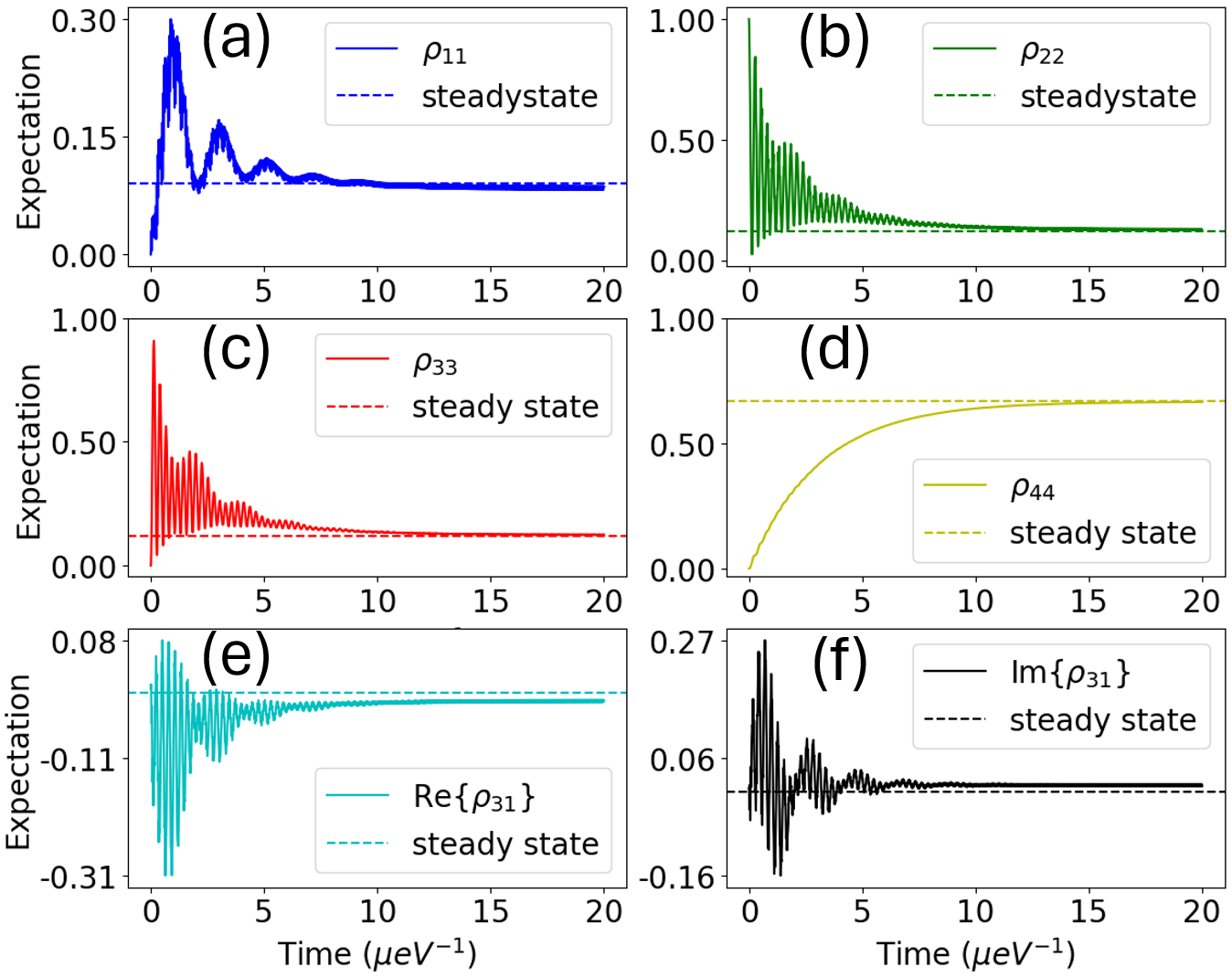}
    \caption{(a)-(d) Dynamics of the density matrix populations in the rotated frame of the two-laser fields with (solid lines) or without (dashed lines) the remaining time-dependent terms. (e)-(f) Real and imaginary parts of the coherence $\rho_{31}$ between $\ket{1}$ and $\ket{3}$. Parameters are as follows: scanning laser power  10\muW, fixed laser power 150\muW,  $\gamma_1 = 0.56$\mueV, $\gamma_3=0.8$\mueV, $\gamma_4 = 0.05$\mueV, $\gamma_\text{nr}=0.06$\mueV, $\gamma_\text{pd}= 0.1$\mueV, $\delta_0= 144$\mueV and $\delta_1 =\delta_3= 0$\mueV}
    \label{Dynamics}
\end{figure}

\noindent To validate this approximation (similar to the usual RWA), we solved the master equation associated to $H'_\text{sys}$ with and without the oscillating terms, for parameters and Rabi frequencies corresponding typically to our experiments, see Fig.~\ref{Dynamics}. Starting from an initial state where the QD system is in the ground state level defined by  $\rho_{22}=1$, we observed for components of the density matrix complete damping of the oscillations on the typical timescale of  10~\textmu eV$^{-1}\simeq7$~ns, and convergence towards the solutions obtained in the  steady-state regime calculated without the oscillating terms. This confirms the validity of the time-independent Hamiltoninan \ref{H2Final} used to interpret the two-laser measurements in Fig.~\ref{Fig4} and Fig.~\ref{Fig5}.
Since the form of Eq. \ref{H2Final} is similar to single laser probing Hamiltonian (Eq. \ref{final_h_rot}), the evaluation of the integrated photoluminescence signal as a function of the scanning laser detuning follows the same procedure as detailed in App.~\ref{AppendixA}.

\newpage
%Create the reference section using BibTeX:
%\bibliography{biblioQDMn}

\end{document}